\documentclass[twocolumn,aps,prl,superscriptaddress]{revtex4}
\usepackage{ae}
\usepackage[T1]{fontenc}
\usepackage[ansinew]{inputenc}
\usepackage{amsmath}
\usepackage{amssymb}
\usepackage{graphicx}
\usepackage{color}
\usepackage[colorlinks]{hyperref}
\usepackage{epstopdf}

\begin{document}

\date{\today}

\title{Spin controlled nanomechanics induced by single-electron tunneling}

\author{D. Radi\'{c}}
\affiliation{Department of Applied Physics, Chalmers University of Technology, SE-412
96 Gothenburg, Sweden}
\affiliation{Department of Physics,
Faculty of Science, University of Zagreb, 10001 Zagreb, Croatia}
\author{A. Nordenfelt}
\affiliation{Department of Physics, University of Gothenburg, SE-412
96 Gothenburg, Sweden}
\author{A. M. Kadigrobov}
\affiliation{Department of Physics, University of Gothenburg, SE-412
96 Gothenburg, Sweden}
\affiliation{Theoretische Physik III,
Ruhr-Universit\"{a}t Bochum, D-44801 Bochum, Germany}
\author{R. I. Shekhter}
\affiliation{Department of Physics, University of Gothenburg, SE-412
96 Gothenburg, Sweden}
\author{M. Jonson}
\affiliation{Department of Physics, University of Gothenburg, SE-412
96 Gothenburg, Sweden}
\affiliation{SUPA, Department of
Physics, Heriot-Watt University, Edinburgh EH14 4AS,
Scotland, UK}
\affiliation{Division of Quantum Phases
and Devices, School of Physics, Konkuk University, Seoul 143-701, Korea}
\author{L. Y. Gorelik}
\affiliation{Department of Applied Physics, Chalmers University of Technology, SE-412
96 Gothenburg, Sweden}


\begin{abstract}
We consider dc-electronic transport through a nanowire suspended between normal- and spin-polarized metal leads in the presence of an external magnetic field. We show that magnetomotive coupling between the electrical current through the nanowire and vibrations of the wire may result in self excitation of mechanical vibrations. The self-excitation mechanism is based  on correlations between the occupancy of the quantized electronic energy levels inside the nanowire and the velocity of the nanowire. We derive conditions for the occurrence of the instability and find stable regimes of mechanical  oscillations.
\end{abstract}

\maketitle

The magnetomotive coupling of electronic and mechanical degrees of freedom induced by an external magnetic field has for many years been a standard tool for implementing electromechanical functionality on the micrometer and nanometer scales \cite{Cleland,VanderZant}. Magnetomotive coupling relies on the electron charge in the sense that an electrical current of charged electrons induces a Lorentz force on the current carrying conductor, while the motion of the conductor itself induces an electromotive force on the charged electrons. However, the electron spin is an additional degree of freedom that may influence electron transport  and thus potentially affect the electromechanics of a device. In particular, it is well known that Zeeman splitting of electronic energy levels affects the current flow in tunnel structures since the electron tunneling rates become spin dependent \cite{Fert,Binsh,Tsukaggoshi,Tsymbal,tsukagoshi2,zhao,Thamankar,merchant}. Furthermore, as sample sizes are reduced to the nano scale, mesoscopic phenomena such as Coulomb blockade of tunneling \cite{Robert} and quantization of electronic energy levels start \cite{Datta} to significantly affect electron transport, leading to a pronounced nonlinear current-voltage characteristics (CVC) \cite{Shi,Souze,ourpaper}. In this Letter we demonstrate that the interplay between spintronics and nanomechanics induced by such mesoscopic effects gives rise to a fundamentally new set of phenomena. In particular, we show below that this interplay may result in self excitation of  mechanical vibrations of a suspended nanowire subjected to an external magnetic field  and a dc voltage bias.

\begin{figure}
\centerline{\includegraphics[width=7.0cm]{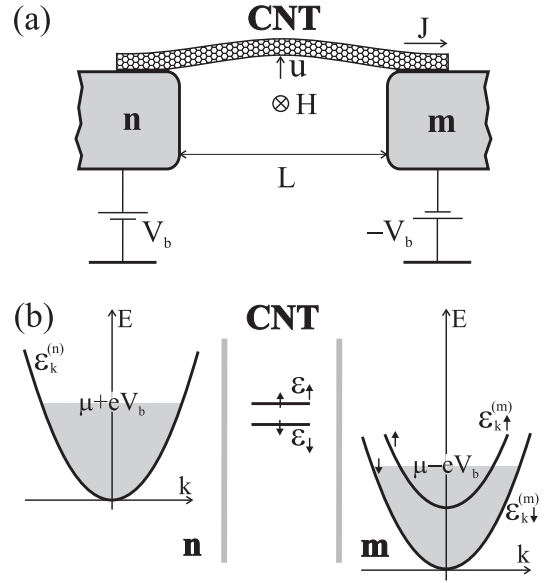}}
\caption{(a) A carbon nanotube (CNT) subject to an external magnetic field  $H$, suspended between normal- (n) and magnetic (m) metal leads biased by voltages $\pm V_b$. (b) Electronic energy scheme for the junction: $\mu$ is the chemical potential, $\varepsilon_{\uparrow,\downarrow}$ are spin-split levels in the CNT,  $\varepsilon_{k}^{(n)}$ and $\varepsilon_{k \sigma}^{(m)}$ are electron energy bands in the leads ($k$ is the wave vector, $\sigma$ is spin).}
\label{junction}
\end{figure}

To be specific we consider the system sketched in Fig.~\ref{junction}a, which shows a single-wall carbon nanotube (CNT) suspended between a normal- and a magnetic metal lead. An external  magnetic field  $H$ is  applied perpendicular to the nanotube and  parallel to the magnetization direction in the magnetic lead. We restrict the nanotube mechanical degrees of freedom to the fundamental bending mode and model it as a classical harmonic oscillator with  frequency $\omega$. The time evolution of the amplitude of the fundamental bending mode $u(\tau)$ is given by the equation
\begin{eqnarray}{\label{mechanics}}
\ddot{u}(\tau)+\frac{\omega}{Q}\dot{u}(\tau)+\omega^2 u(\tau)&=& \Im(\tau), \\
\Im(\tau)  &=& -m^{-1}\mathrm{Tr}\left(\partial_{u}H_{e}\rho(\tau)\right)\,,\nonumber
\end{eqnarray}
where $Q$ is the quality factor characterizing the dissipative processes induced by the coupling to heat baths in the leads, $m$ is the mass of the oscillator, $\rho$ is the density matrix of the electronic subsystem described by the Hamiltonian $H_{e}$, defined as
\begin{eqnarray}{\label{Hamiltonian}}
H_{e}&=&H_{l}+H_{d}+H_{t}\\
H_{l}&=& \sum_{k,\sigma}\varepsilon_{k}^{(n)}  a_{k \sigma}^{\dagger} a_{k \sigma}+ \sum_{k,\sigma} \varepsilon_{k \sigma}^{(m)}  c_{k \sigma}^{\dagger} c_{k \sigma} \nonumber \\
H_{d}&=& \sum_{\sigma} \varepsilon_{\sigma} d_{\sigma}^{\dagger} d_{\sigma}+
Ud_{\uparrow}^{\dagger}d_{\downarrow}^{\dagger}d_{\downarrow}d_{\uparrow} \nonumber\\
H_{t}(\tau)&=&t_{n}e^{i (eV_{b}\tau -u(\tau)\mathrm{p}_{L}\tau)/\hbar}\sum_{k,\sigma} a_{k \sigma}^{\dagger} d_{\sigma}\nonumber\\
&+&t_{m}e^{-i (eV_{b}\tau -u(\tau)\mathrm{p}_{L}\tau)/\hbar}\sum_{k,\sigma} c_{k \sigma}^{\dagger} d_{\sigma}+ h.c.\,.\nonumber
\end{eqnarray}
Here $a_{k \sigma}^{\dagger}$ and $c_{k \sigma}^{\dagger}$ are creation operators for electrons in the normal- and  spin-polarized metal leads, respectively, $d_{\sigma}^{\dagger}$ is the creation operator for electrons in the nanotube, while the index $\sigma=\uparrow,\downarrow$ labels the electronic spin.

The Hamiltonian $H_l$ in Eq.~(\ref{Hamiltonian}) describes the two leads as reservoirs of non-interacting electrons, while the expression for $H_{d}$ describes the electronic properties of the nanotube considered as a quantum dot, with its second term representing the Coulomb interaction between two electrons with different spin projections. The spacing between electronic levels inside the nanotube is assumed to be large enough with respect to the symmetrically applied bias voltage $2V_{b}$ for only a single pair of Zeeman-split electron energy levels, $\varepsilon_{\uparrow,\downarrow}=\epsilon_0 \pm \mu_B H /2$, to be relevant for electronic transport
(if $H=0$ the energy level $\epsilon_0$ is spin-degenerate;
$\mu_B$ is the Bohr magneton).

The Hamiltonian $H_{t}$ in Eq.~(\ref{Hamiltonian}) describes electron tunneling between the nanotube and the leads in terms of the tunneling amplitudes $t_n$ and $t_m$  between the nanotube and the normal and magnetic lead, respectively. The phases of the tunneling amplitudes depend on the nanotube deflection \cite{AA}, the parameter $\mathrm{p}_L=\alpha eHL/c$ gives  the nanotube momentum change induced by the Lorentz force when an electron tunnels from the nanotube to a lead, and $\alpha \sim 1$ is a numerical factor determined by the spacial profile of the fundamental mode.

The density matrix $\rho$ obeys the Liouville-von Neumann equation. Introducing the interaction representation, $\rho(\tau)=e^{-iH_{0} \tau/\hbar}\tilde{\rho}(\tau)e^{iH_{0}\tau/\hbar}$ ; $H_{0}=H_{l}+H_{d}$, and exploiting the reduced density matrix Ansatz \cite{CP} $\tilde{\rho}(\tau)=\rho_{d}(\tau)\cdot\rho_{l}$ one gets the following equations for  $\rho_{d}$ and the force term, $\Im$, in Eq.~(\ref{mechanics}):
\begin{eqnarray}{\label{rd,F}}
 \Im(\tau)=-\frac{1}{i \hbar m}\mathrm{Tr}
 \left\{\int_{-\infty}^{\tau}d\tau'\left[\partial_{u}\tilde{H}_{t}(\tau),\tilde{H}_{t}(\tau')\right]\rho_{d}(\tau')\rho_{l}\right\},\nonumber\\
 \dot{\rho}_{d}(\tau)=-\frac{1}{\hbar^{2}} \mathrm{Tr}_{l}\left\{\int_{-\infty}^{\tau}d\tau'\left[\tilde{H}_{t}(\tau),\left[\tilde{H}_{t}(\tau'),\rho_{d}(\tau')\rho_{l}\right]\right]\right\}.
\end{eqnarray}
Here $\tilde{H}_{t}(\tau)=e^{iH_{0}\tau/\hbar}H_{t}(\tau)e^{-iH_{0}\tau/\hbar}$ and  $\rho_{d}(\tau)=\mathrm{Tr}_{l}\{\tilde{\rho}\}$, where $\mathrm{Tr}_{l}$ denotes a trace over all electronic states in the leads, while  $\rho_{l}=e^{-(H_{l}-\mu N)/k_{B}T}/\mathrm{Tr}_{l}\{e^{-(H_{l}-\mu N)/k_{B}T}\}$, $\mu$ is the chemical potential, and $N$ is the electron number operator in the leads.

The two equations (\ref{rd,F}) describe the coupled dynamics of the electronic and mechanical subsystems. To demonstrate the interplay between spintronics and nanomechanics  we  analyze them for the case when the bias voltage is applied in such a way that the normal lead  serves as a source (S) electrode while the magnetic lead acts as a drain (D) (see Fig.~1b).

Two additional conditions are assumed  to be satisfied:
First, that $U \gg eV_b\sim \tilde{\varepsilon}_{\sigma} \equiv \varepsilon_\sigma -\mu \gg k_{B}T$, which means that the charge transfer from the drain electrode to the nanotube  is exponentially suppressed, while the  strong electron-electron interaction inside the nanotube imposes the Coulomb blockade constraint $Tr(d^{\dag}_{\uparrow}d^{\dag}_{\downarrow}d_{\uparrow}d_{\downarrow} \rho_{d})=0$. Secondly, that $\omega \ll \Gamma_{S(D)}$, $k_{B}T/\hbar$. Here $\Gamma_{S(D)}^{\sigma}=2\pi\hbar^{-1}t^{2}_{n(m)}\nu_{n(m)}^{{\sigma}}$, and the electronic density of states $\nu_{n(m)}^{\sigma}$ are assumed to be independent of energy for both leads and also independent of spin in the source lead, so that $\nu_{n}^{{\uparrow}}=\nu_{n}^{{\downarrow}}$ and hence $\Gamma_{S}=\Gamma_{S}^{\uparrow,\downarrow}$.

In the case of a low mechanical oscillator frequency one can use a quasi-static approximation to  solve Eqs. (\ref{mechanics}) - (\ref{rd,F}). Neglecting the time dependence of $\rho_{d}$ and using the approximation $u(\tau)-u(\tau')\simeq  \dot{u}(\tau)(\tau-\tau')$ one finds that
\begin{eqnarray}
  \Im(\tau)&=& \frac{\alpha HL}{m} J(V_{b}-\alpha HL \dot{u}){\label{Force}}, \\
  J(V) &=& \frac{e\Gamma_{S}\Gamma_{D}^{\downarrow}}{\Gamma_{S}+ \Gamma_{D}^{\downarrow} +\Gamma_{S}\exp\{\left(eV-\tilde{\varepsilon}_{\uparrow} \right)/k_{B}T\} }\,.
\label{current1}
\end{eqnarray}
To arrive at Eq.~(\ref{current1}) we assumed the magnetic lead to be fully spin polarized, so that $\nu_{m}^{{\uparrow}}=0$ \cite{Shi,Souze,ourpaper}. To assume partial polarization would be more realistic, but not important for our conclusions, as will be discussed below.

The physical interpretation of this result is obvious: the force acting on the nanotube is the Lorentz force induced by the quasi-stationary current $J(V)$, while the effective voltage $V$ is comprised of the  bias voltage $V_{b}$ and the electromotive force $\alpha HL \dot{u}$ that is induced by the motion of the nanowire in the external magnetic field. Linearizing Eq.~(\ref{mechanics}) with $\Im $ given by Eq.~(\ref{Force}) one finds that the electromotive coupling induces an effective damping or pumping of mechanical vibrations in accordance with the sign of the differential conductance $J^{'}= dJ/dV$. In the case of pumping, which corresponds to $J^{'}<0$, one can expect an electromechanical instability resulting in self-excited oscillations of the nanowire if $Q$ is large enough and the instability parameter $\beta$ is positive, i.e. if
\begin{equation}
 \beta \equiv - \frac{\alpha^{2}H^{2}L^{2}}{ m}J^{'}(V_b)-\frac{\omega}{Q}>0\,.
\label{instabilitycriteria}
\end{equation}
It is obvious from Eq.~(\ref{current1}) that the differential conductance $J^{'}(V)$ is negative for voltages such that $|V-\tilde{\varepsilon}_\uparrow/e| \lesssim k_BT/e$. Even if the magnetic lead is less than 100\% polarized, as assumed in  (\ref{current1}), $J^{'}(V)$ would still be negative in the same voltage interval \cite{Shi,Souze,ourpaper}. Therefore, the necessary condition $J^{'}(V)<0$ for an instability to occur does not require 100\% polarization. Nevertheless, for the sake of clarity, we assume this to be the case in what follows.

The  mechanism of the instability is most transparent when $eV_{b}=\tilde{\varepsilon}_{\uparrow}$, as in
Fig.~\ref{junction}b, and $\mu_B H \gg k_{B}T$. Under such conditions  electrons that jump from the S-lead to the spin-up level in the nanotube can only jump back, since they can not reach any spin-up state in the D-lead. Such  processes lead to random changes of the nanotube momentum by $\pm p_{L}$, which results in  a diffusive motion in its phase space. However, an electron that tunnels to the spin-down level under the same circumstances can only proceed by tunneling to the D-lead. This is because there are no empty states in the S-lead at accessible energies. Tunneling through the spin-down channel changes the net nanotube momentum by $p_{L}$ and hence its mechanical energy by $p_{L}\dot{u}(\tau)$. Therefore, correlations between the probability for tunneling through the spin-down level and the direction of the nanotube motion can give rise to either damping or pumping of its mechanical vibrations. Precisely such correlations appear due to the  Coulomb blockade of tunneling to the spin-down  level  when the spin-up level is occupied. Indeed, the population of the spin-up level increases when the nanotube moves in such a direction that the electromotive force shifts the ``effective" chemical potential, $\mu_{\rm eff}=\mu - \alpha HL \dot{u}$, up. A shift in this direction means that electron tunneling to the spin-down level is decreased due to the Coulomb blockade. When the nanotube moves in the opposite direction the situation is reversed and the probability for electrons to tunnel to the spin-down level is enhanced. This means that, on the average, momentum transfer to the nanotube is more probable if the  nanowire moves in this particular direction. In short, the spin-up level in the nanotube  serves as an effective gate, which --- depending on the direction in which the nanotube moves --- increases or decreases the probability for electron tunneling from the source- to the drain electrode via the spin-down level; the only available channel in our case.
The described correlations between the tunneling processes and the nanotube velocity lead to  a ``pumping" of energy into the mechanical subsystem. If the rate of pumping is larger than the rate of dissipation (due to damping), the result is a mechanical instability.

To analyze how the instability described above evolves we impose the Ansatz $u(t)=u_0 + A(t)cos(\omega t)$. Here $u_0$ is the stationary shift of the nanowire position induced by the average Lorentz force and $A(t)$ is the amplitude of the harmonic oscillations, which by assumption is slowly varying, so that $\dot{A} \ll \omega A$. Inserting this Ansatz into  Eq.~(\ref{mechanics}), using Eqs.~(\ref{Force}) and (\ref{current1}) and averaging over the fast oscillations one gets an  equation for the amplitude of the form
\begin{eqnarray}
\label{ampcond}
\dot{A} = -\Phi(A);\\
\Phi(A) = \frac{\omega A}{2Q} &+& \frac{\alpha H L}{m \omega}
 \int_{-\pi}^{\pi} \frac{d\phi}{2\pi} J(V_b + \alpha H L \omega A \sin\phi) \sin\phi. \nonumber
\end{eqnarray}
\begin{figure}
\centerline{\includegraphics[width=8.0cm]{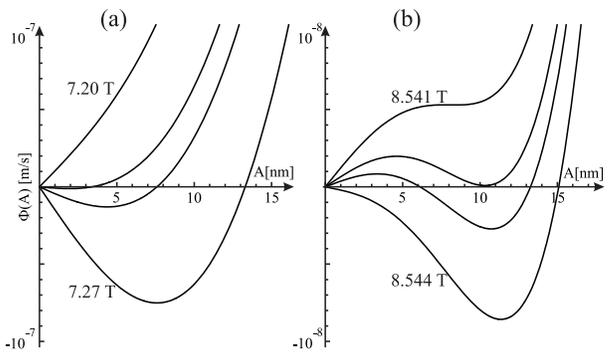}}
\caption{Behavior of the function $\Phi(A)$ defined in Eq.~(\ref{ampcond}) for different magnetic fields leading to (a) soft excitation of nanowire oscillations at  $\Delta V\equiv  V_b-(\epsilon_0 -\mu)/e= 0.34$~mV and (b) hard excitation at $\Delta V= 0.41$~mV. Other parameter values are $T=0.2$~K, $Q=3 \cdot 10^4$, $\Gamma_{D}^{\downarrow}/\Gamma_{S} = 0.4$, $\omega / \Gamma_S = 0.2$}
\label{satamp}
\end{figure}
The first  term in this expression for $\Phi(A)$ describes the damping of mechanical oscillations due to a coupling to the thermal environment, while the second term  describes  the pumping  generated by the electronic current. If the condition (\ref{instabilitycriteria}) is satisfied, pumping dominates over damping at small vibration amplitudes $A$ and hence the amplitude increases with time according to Eq.~(\ref{ampcond}). However, as can be inferred from Eq.~(\ref{current1}), at large amplitudes $A\gg k_BT/eH L \omega$ the pumping term in Eq.~(\ref{ampcond}) saturates at $\sim (HL/m\omega)J_{0}$, where $J_{0}\sim e\Gamma_{D}^{\downarrow}$ is the characteristic current through the system. Accordingly,  the damping and pumping terms cancel each other at a finite amplitude $A_{st}\sim (QHL/m\omega^2)J_{0}$.

The exact value of the amplitude of stationary oscillations can be obtained by solving the equation $\Phi(A_{st})=0$. A numerical analysis of the function $\Phi (A)$ shows that, depending on  the bias voltage,   the onset of stationary nanowire oscillations in an increasing magnetic field can be either soft ($A_{st}$ increases continuously from zero) or hard ($A_{st}$ jumps to a finite value), see  Fig.~\ref{satamp}. In order to investigate this situation analytically we assume the instability parameter to be small, $|\beta(V_b,H)|\ll 1$, and expand $\Phi (A)$ in a Taylor series. Keeping terms up to the third order in $A$ one finds
\begin{equation}
\label{Phyexpansion}
\Phi(A) \approx -\frac{A}{2}\Bigl( \beta -\frac{(\alpha HL)^4\omega^2}{8 m} J^{'''}(V_b)A^2\Bigr)\,.
\end{equation}
\begin{figure}
\centerline{\includegraphics[width=8.0cm]{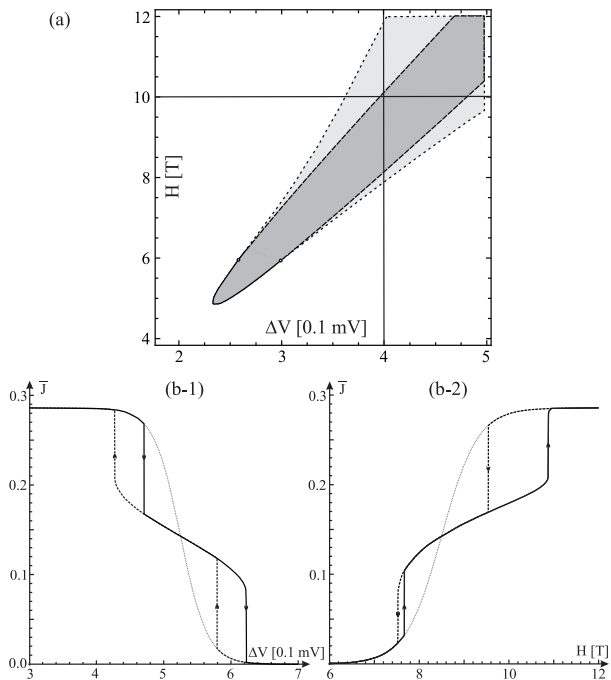}}
\caption{(a) Stability diagram for nanowire oscillations in the ($H, \Delta V$) plane, where $\Delta V =V_b-(\epsilon_0 -\mu)/e$. In the white and dark grey regions there is only one stable stationary state corresponding to a non-moving and a vibrating nanowire, respectively. The light gray region is where both these states are stable.
(b) Average current in units of $e\Gamma_S$, in (b-1) as a function of  bias voltage $\Delta V$ for $H=10$~T [horizontal line in (a)] and in (b-2) as a function of magnetic field $H$ for  $\Delta V =0.4$~mV [vertical line in (a)]. The full (dashed) lines show the result for an ascending (descending) magnetic field. The gray dotted curves correspond to a static nanowire. Other parameters: $T=0.2$~K, $Q=3 \cdot 10^4$, $\Gamma_{D}^{\downarrow}/\Gamma_{S} = 0.4$, $\omega / \Gamma_S = 0.1$. }
\label{instdiagram}
\end{figure}
a) \emph{Soft excitation of nanowire oscillations}, Fig. \ref{satamp}a.  It is readily seen from Eq. (\ref{Phyexpansion}) that the onset of nanowire oscillations is soft in the interval  $V^*_1 < V_b < V^*_2$, where $J^{'''}(V_b) > 0$ (here $V^*_{1,2}$ are the points of inflection of the negative differential resistance curve $J^{'} (V)$); for $\beta(V_b,H)> 0$ the non-moving state of the wire is unstable and stationary oscillations with amplitude $A_{st}^{(1)} \propto \sqrt{\beta/J^{'''}}$  appear spontaneously, smoothly increasing in amplitude with increasing values of $\beta(V_b, H)$ (that is with increasing $H$).

b) \emph{Hard excitation of nanowire oscillations}, Fig. \ref{satamp}b.
With bias voltages $V_b$ for which $J^{'''}(V_b)< 0$ (i.e., $V_b<V^*_1, V_b>V^*_2$),  the instability develops in a qualitatively different way.
There are now two bifurcation points: $H_{c}^{(1)}(V_b)$, at which  the curve $\Phi(A,H)$ ``touches" the A-axis from above (see Fig.~\ref{satamp}b)
and $H_{c}^{(2)}(V_b)>H_{c}^{(1)}(V_b) $ at which $\beta(H_{c}^{(2)}(V_b)) =0$. As long as $H<H_{c}^{(1)}(V_b) $ the non-moving state of the nanowire is stable, as shown in Fig.~\ref{instdiagram}. However, in the interval $H_{c}^{(1)}(V_b)<H<H_{c}^{(2)}(V_b) $ the system is  bistable since here both the non-moving state ($A=0$) and nanowire vibrations of a finite amplitude $A_{st}\sim (QHL/m\omega^2)J_{0}$ (see above) are stable.
When $H>H_{c}^{(2)}(V_b) $ the non-moving state is unstable and  vibrations with a finite amplitude  is the only stable stationary  state of the  nanowire.

The existence of a region of magnetic fields and bias voltages for which the system is  bistable with respect to the vibration amplitude results in a hysteretic behavior of the averaged current $\bar{J}=\int J(V_b + \alpha H L \omega A \sin\phi)d\phi/2\pi$ under a change of magnetic field or bias voltage. Figure~\ref{instdiagram} shows that the width of the hysteresis loop is $\sim 0.1$~mV with respect to the bias voltage and $\sim 1$~T with respect to the magnetic field, while its height is $\sim J_0$.

In conclusion, we have considered the mechanical properties of a nanowire suspended between  two metal leads, one nonmagnetic and one magnetic.  We have shown that interplay between Coulomb blockade of tunneling and spin-dependent single-electron tunneling  gives rise to a fundamentally new response to a magnetomotive coupling of the electrical current through the nanowire and mechanical oscillations of the nanowire. In particular, we have demonstrated that in the presence of an external static magnetic field, mechanical vibrations of a dc voltage-biased nanowire can be self-excited. In contrast to the resonant excitation of nanowire oscillations that may be induced by an electrical or magnetic ac signal, the  amplitude of these self-excited stationary vibrations is not limited by any resonant condition and, as a result, they can be large.
In a realistic experimental situation with a CNT resonator of length $L\sim 1$~$\mu$m, vibration frequency $\omega\sim 2\pi\times 200$~MHz, and quality factor $Q\sim 10^4$ carrying a characteristic current $J_0 \sim 1$~nA,  the vibration amplitude is $A_{st}\sim 10$~nm at $T\sim 0.1$~K in a magnetic field of $\sim 10$~T. In principle, oscillations with such  an amplitude could be directly monitored by clever imaging techniques \cite{Lexholm}.
We have also demonstrated that the onset and disappearance of these mechanical vibrations are manifest in a pronounced hysteretic behavior of the averaged electrical current through the structure. This hysteresis also facilitates detection of the self-excited nanowire vibrations in an experiment. In general, spin accumulation at the nanowire-magnetic lead contact may also result in an instability or hysteretic behavior of the current-voltage characteristic due to the spin transfer torque effect \cite{Slonczewski}. However, this requires \cite{Ralf} current densities $\sim 10~\mu$A/(nm)$^2$, which are orders of magnitude  higher than those considered here.

{\it Acknowledgement.} Financial support from  the Swedish VR and SSF, the European Commission (FP7-ICT-2007-C; projects no 233952 QNEMS and no 225955 STELE), and the Korean WCU program funded by MEST/NFR (R31-2008-000-10057-0) is gratefully acknowledged.

\end{document}